\documentclass{llncs}

\usepackage{graphicx}
\usepackage{hyperref}

\begin{document}

\title{Formal Verification of a Geometry Algorithm: \\
A Quest for Abstract Views and Symmetry \\ in Coq Proofs}

\author{Yves Bertot\orcidID{0000-0001-5052-3019}}

\institute{Inria Sophia Antipolis -- M\'editerran\'ee, \\
2004 route des Lucioles, 06902 Sophia Antipolis Cedex, France \\
Universit\'e C\^ote d'Azur\\
\email{yves.bertot@inria.fr}}

\maketitle

\begin{abstract}
  This extended abstract is about an effort to build a formal description of a
  triangulation algorithm starting with a naive description of the
  algorithm where triangles, edges, and triangulations are simply
  given as sets and the most complex notions are those of boundary and
  separating edges. When performing proofs about this algorithm,
  questions of symmetry appear and this exposition attempts to give an
  account of how these symmetries can be handled. All this work relies
  on formal developments made with Coq and the mathematical components
  library.
\end{abstract}

\section{Introduction}

Over the years, proof assistants in higher-order logic have been advocated as tools to improve the
quality of software, with a wide range of spectacular results, ranging from compilers, operating systems,
distributed systems, and security and cryptography primitives.  There are now good reasons to believe
that any kind of software could benefit from a formal verification using a proof assistant.

Embedded software in robots or autonomous vehicles has to maintain a view of the geometry of the
world around the device.  We expect this software to rely on computational geometry.  The work described
in this extended abstract concentrates on an effort to provide a correctness proof for algorithms that construct triangulations.

\section{An Abstract Description of Triangulation}
Given a set of points, a triangulating algorithm returns a collection of triangles that must cover the space between
these points (the convex hull), have no overlap, and such that all the points of the input set are vertices of at least
one triangle.  When the input points represent obstacles, the triangulation can help construct safe routes between these obstacles, thanks to Delaunay triangulations and Vorono\"{\i} diagrams.

The formal verification work starts by providing a naive and abstract view of the algorithm that is later refined into
a more efficient version.   Mathematical properties are proved for the naive version and then modified for successive
refinements.  When the proof is about geometry and connected points, it is natural to expect symmetry properties
to play a central role in the proofs.  In this experiment, we start with a view of triangles simply as 3-point sets.  
We expect to refine this setting later into a more precise graph structure, where each triangle is also equipped with a link to its
neighbors and costly operations over the whole set of triangles are replaced by low constant time operations that
exploit information that is cached in memory.

From the point of view of formal verification, the properties that
need to be verified for the naive version are the following ones: all triangles have the right number of elements,
all points inside the convex hull are in a triangle, the union of all the triangles is exactly the input, and there is no
overlap between two triangles.

The naive algorithm relies on the notion of separating edges of a triangle with respect to
a point:  for a triangle \(\{a, b, c\}\) and a fourth point \(d\), the point \(c\) is {\em separated} from \(d\) if \(c\) and \(d\) appear on
different sides of the edge \(\{a, b\}\).  At this point, it appears that life is much easier if we take the simplifying assumption
that three points of the input are never aligned.  This assumption is often taken in the early literature on computational
geometry and we will also take it.

The point \(d\) is inside the triangle \(\{a, b, c\}\) exactly when no element of the triangle is separated from the point \(d\).
When the point \(d\) is outside the triangle, for instance when \(c\) is separated from \(d\), the edge \(\{a,b\}\) will be called {\em red}.  An edge that is not red will be called {\em blue}.

Another important notion is the notion of boundary edge.  An {\em edge} of the triangulation is a 2-point subset of
one of the triangles in the triangulation, a {\em boundary edge} is an edge that belong to exactly one triangle.  Boundary edges are triangle edges, and as such they can be blue or red
with respect to a new point.

The algorithm then boils in the following few lines:

Take three points from the input: they constitute the first triangle, then take the points one by one.
\begin{itemize}
\item If the new point is inside an existing triangle, then remove this triangle from the triangulation and then add the
three triangles produced by combining the new point and all edges of the removed triangle.
\item If the new point is outside, then add all triangles obtained by combining the new point with all red boundary edges.
\end{itemize}
This algorithm terminates when all points from the input have been consumed.

\section{Specifying the Correctness of the Algorithm}
This algorithm is so simple that it seems proving it correct should be extremely simple.  However, geometry properties
play a significant role, as is already visible in the specification.

That the triangulation only contains 3-set seems obvious, as soon as the input set does contain three points.  When
there are more than 3 points, say \(n\) points, we can assume by induction that the triangulation of the first n-1 points
contains only-3 sets.  Then, whether the new point is inside an existing triangle or outside, the new elements of the
triangulation are obtained by adding the new point to edges of the previous triangulation.  These operation always
yield 3-point sets.

To verify that the union of all triangles is the input set, we need to show that at least one triangle is created when
including a new point.  This is surprising difficult, because it relies on geometry properties.   If the new point is
inside an existing triangle, the algorithm obviously includes in the triangulation three triangles that contain the new point.
However, when the point is not inside a triangle, there is no simple logical reason for which there should exist a boundary
edge that is also red.  This requires an extra proof with geometrical content.  Such a proof was already formally verified  by Pichardie and Bertot \cite{PichardieBertot01short}.

With respect to boundary edges, when the triangulation is well-formed, all boundary edges
should form the convex hull of the input set.  In other words, for every point inside the
convex hull, all boundary edges should be blue.
\section{Formal Proof}
When performing the proofs, it is interesting to exploit all the symmetry that can be found.  In
paper proofs, it is often enough to explicit one configuration and state rapidly that many other
configurations can be proved similarly by symmetry.

\subsection{Combinatorial Symmetries of Triangles}
One example is the natural symmetry of triangles.  When considering triangles, Knuth \cite{KnuthAxiomsHulls} proposed that they should be viewed as ordered triplets \(abc\), such that one turns left when following the edges from \(a\), \(b\) and then to \(c\).  Of course, if one views triangles simply as sets, it does not make sense to distinguish between oriented and non-oriented triangles.  Thus, we need to add structure to the set, which we do by giving names to the elements.  Now, when giving these names, we can do it in a way that ensures the obtained triangle to be oriented.  When doing our formalization work, it becomes natural to name \(t_1\), \(t_2\), \(t_3\) the three points of \(t\).

In practice, we don't use integers for indexing the elements, because this means we would have to give a meaning
to \(t_{18}\).  Instead, we use the type of integers smaller than 3 and we use the fact that this set can be given the structure of a
group.  The mathematical component library already provides such a structure, noted {\tt 'I\_3}.  We profit from it and call \(0\), \(1\), and \(-1\)
the three elements.  A characteristic property in our development will be that \(t_i\), \(t_{i+1}\), \(t_{i-1}\) form an oriented
triangle, of course with the convention that \(i + 3 = i\) and \(0 -1 = 2\) when dealing with elements of {\tt 'I\_3}.  This is a first way in which we attempt to deal with symmetry.  This
is supported by the finite group concepts in the library.

We define a function {\tt three\_points} that maps any set of type {\tt \{set P\}} (this is the mathematical components' notation for sets of elements of {\tt P}) to a function from
{\tt 'I\_3} to {\tt P}.  This function is defined in such a way that it is injective and its image is included in its first argument as soon as this set has at least three points and the images
of \(0\), \(1\), and \(-1\) form an oriented triplet.

\subsection{Geometric Symmetries of Triangles}
Other symmetries come up when considering oriented triangles in the plane.  In his study of convex hull algorithms \cite{KnuthAxiomsHulls}, Knuth expresses that the following 5 properties
are to be expected from the orientation predicate, when the 5 points \(a\), \(b\), \(c\), \(d\), and \(e\) are taken to be pairwise distinct, and noting simply \(abc\) to express that one
turns left when following the path from \(a\) to \(b\) and then \(c\).
\begin{enumerate}
\item \(abc \Rightarrow bca\)
\item \(abc \Rightarrow \neg bac\)
\item \(abc \vee bac\)
\item \(abd\wedge bcd \wedge cad \Rightarrow abc\)
\item \(abc\wedge abd \wedge abe \wedge acd \wedge ade \Rightarrow ace\)
\end{enumerate}
Knuth calls these properties axioms of the orientation predicate and we will follow his steps, even though from the logical point of view, these properties are not really axioms because we
can prove them for a suitable definition of the orientation predicate (using the points' coordinates and determinants).
 
The first axiom essentially says that from the geometrical point of view, triangles exhibit a ternary symmetry.  The second one makes it slightly more precise by expressing that not
any order sequence of three points forms an oriented triangle.  The third one states that
we are working under the assumption that no three points in the data set are aligned.  The fourth axiom expresses that the combination of three adjacent oriented triangles lead
to fourth one.  It also has a natural ternary geometric symmetry, which is perhaps easier
to see in the following drawing:
\begin{center}
\includegraphics[scale=0.5,trim={3cm 16cm 7cm 7cm}, clip]{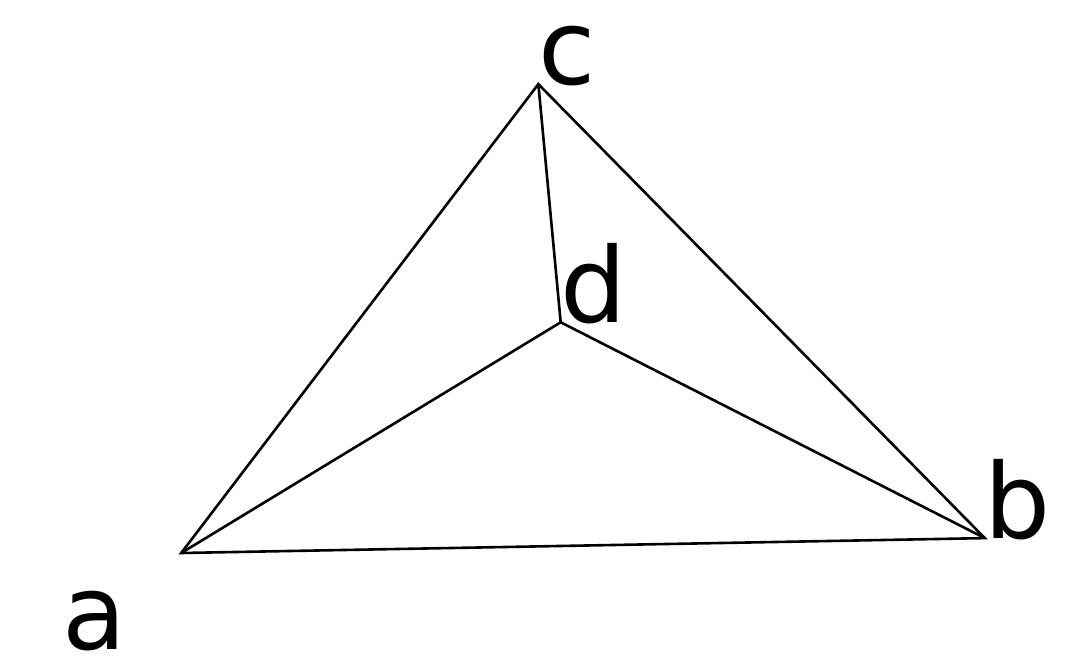}
\end{center}

Axiom 5 describes relations of four points relative to a pivot, in this case \(a\).  It can
be summarized by the following figure, where the topmost arrow (in blue) is a consequence
of all others.

\begin{center}
\includegraphics[scale=0.7,trim={4cm 16cm 4cm 4cm}, clip]{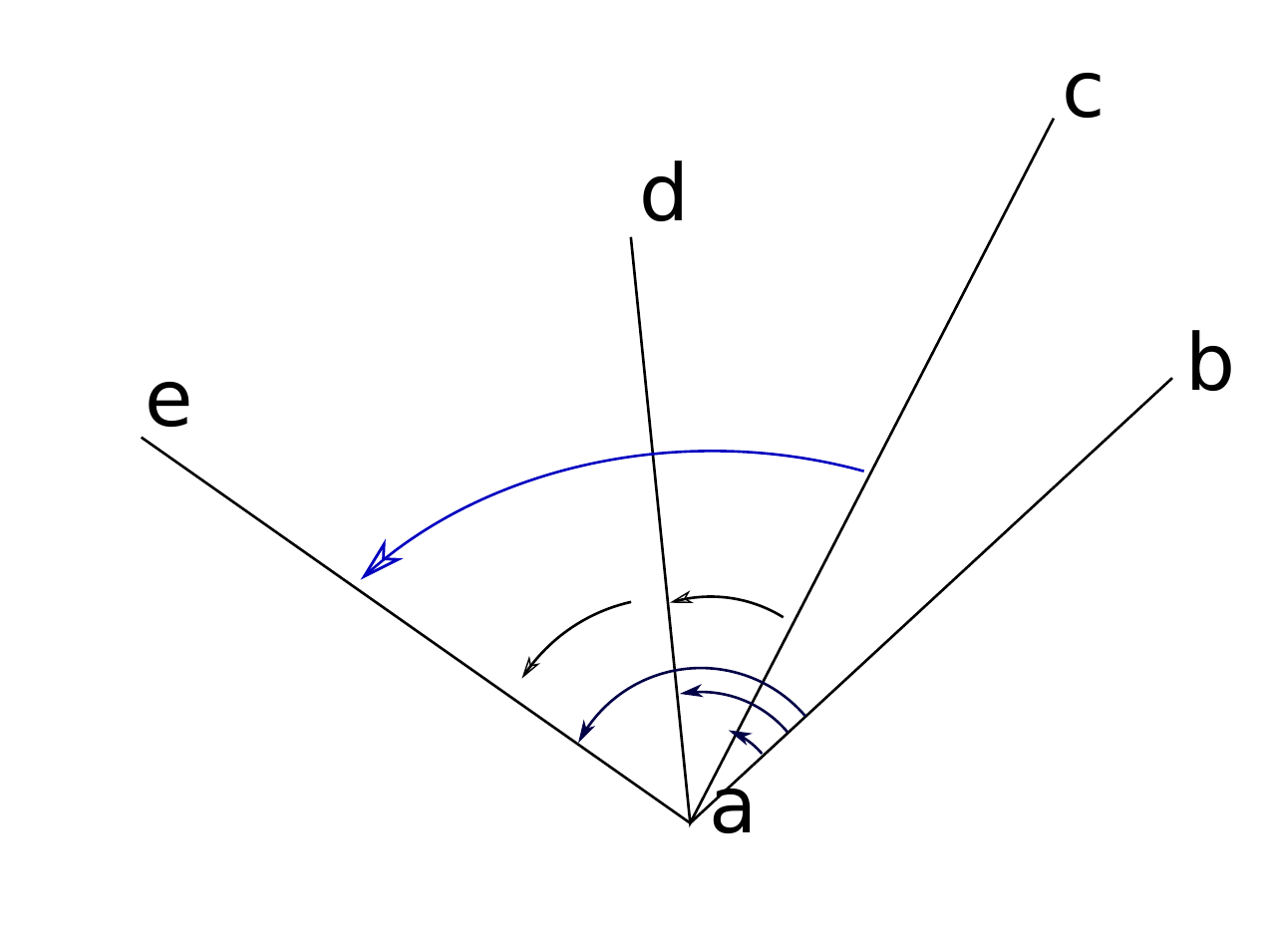}
\end{center}
To have a symmetric collection of axioms, we would actually need a similar statement, but with all points pivoting around \(b\).  Knuth also recognizes this need and actually shows
that the symmetric picture (an axial symmetry) is a logical consequence of all other axioms.

Using these axioms, we should be able to prove a statement like the following
{\em if all vertices of a triangle \(\{c,d,e\}\) lay on the left of a segment \([a,b]\), then any point \(f\) inside the triangle also lays on the left of the segment.   This should also
be true when one or both of \(a\) and \(b\) is element of \(\{c,d,e\}\).}

A human readable form of this proof works by first studying the case the where sets \(\{a, b\}\) and
\(\{c,d,e\}\) are disjoint, noting that there should be at least one
edge of the triangle that is red with respect to both \(a\)  and by supposing,
without loss of generality that this edge is \([c,d]\).    This proof already relies on 9
uses of Knuth's fifth axiom or its symmetric.

For a human reader, the exercise of renaming points is easily done, but for a computer, 
the three points \(c\), \(d\), and \(e\) are not interchangeable and performing the ``without loss of generality'' step requires a technical discussion with  three cases to consider,
where Knuth's fifth axiom is used once again.  In total, if no step was taken to exploit
the symmetry, this means that the proof would require 28 uses of Knuth's fifth axiom and
since this proof has 5 premises, this corresponds to a proof complexity that it really cumbersome for the human mind.

More uses of symmetry have to be summoned to treat the cases when \(a\) and \(b\) may appear among the vertices \(c\), \(d\), and \(e\), depending on whether it is \(a\), \(b\), or both that belongs to the triangle when \(c\), \(d\), or \(e\) are all on the left of the \([a,b]\) segment.

\subsection{Symmetries with respect to the Convex Hull}
In two dimensions, the boundary edges of the convex hull form a loop where no edge plays a more significant role than the other.  It is natural to think that the ternary symmetry of triangles should generalize to such a loop, but with the added ingredient that the size of the loop is an arbitrary number \(n\), larger than 3.  To cope with this source of symmetry, we did
not choose to exhibit a mapping from {\tt 'I\_n} to the type of points, but rather to
indicate that there exists a function \(f\), such that \([x; f(x)]\) is always a boundary
edge when \(x\) is taken from the union of the boundary edges, of the triangulation and all the other points of the triangulation are always on the left side of the segment \([x; f(x)]\).

To handle this point of view, the mathematical components library provides a notion of
orbit of a point for a function.

When one considers the operation of adding a new point outside the convex hull, it is not true
anymore that all boundary edges are equivalent.  Some edges are red, some edges are blue.  In fact,
it is possible to show that all red boundary edges are connected together, so that there are exactly two points,
which we can call the {\em purple} points that belong to two edges of different color.  The role
of these two points is symmetric, but they can be distinguished: for one of them, which we call \(p_1\),
the edge \([p_1, f(p_1)]\) is a red  boundary edge and \([f^{n-1}(p_1),p_1]\) is a blue boundary edge, for the
other, which we call \(p_2\), the edge \([p_2, f(p_2)]\) is blue and \([f^{n-1}(p_2),p_2]\) is red. 
In fact, there exists a number \(n_r\) such that \(f^{n_r}(p_1) = p_2\),
all segments \([f^{k}(p_1), f^{k+1}(p_1)]\) are red boundary edges when \(0 \leq k < n_r\) and
all segments \([f^{k}(p_1), f^{k+1}(p_1)\) are blue when \(0 \leq n_r < n\).

In principle, all statements made about \(p_1\) are valid for \(p_2\), mutatis mutandi.  In practice, performing the proofs of the symmetric statement formally often relies on copying and pasting the proofs obtained for the first case, and guessing the right way to exploit
the known symmetries, for example by replacing uses of Knuth's fifth axiom by its symmetric.  The alternative is to make the proof only once and make the symmetry explicit, but the last step is often as difficult as the first one.

The existence of a cycle for the function \(f\), so that \(f^{k+n} = f^k\) also plays a role in
the proof.   Reasoning
modulo \(n\) appears at several places during the proof, but for now we have not found a satisfactory way to exploit this fact.

\section{Related Work}
The formal verification of computational geometry algorithms is quite rare.  A first attempt with convex hulls was provided by Pichardie and Bertot \cite{PichardieBertot01short} where the only data structure used was that of lists but the question of non general positions (where points may be aligned) was also studied.  Notable work is provided by Dufourd and his colleagues \cite{DBLP:journals/tcs/DehlingerD04,DBLP:journals/jar/Dufourd09,DBLP:conf/itp/DufourdB10,DBLP:journals/comgeo/BrunDM12}.  In particular, Dufourd advocated the use of hypermaps to represent many of the data-structures of computational geometry.  In this work,
we prefer to start with a much more naive data structure, closer to the mathematical perspective, which consists only of viewing the triangulation as a set of sets.  Of course,
when considering optimisations of the algorithm, where some data is pre-computed and
cached in memory, it becomes useful to have more complex data-structure, but we
believe that the correspondence between the naive algorithm and the clever algorithm can
be described as a form of refinement which provides good structuring principles for the whole
study and for the formal proof.  In the end, the refinement will probably converge towards the data-structure advocated by Dufourd and his colleagues.
It should be noted that the hypermap data-structure was also used by Gonthier in his study 
of the four-color theorem \cite{Gonthier4CT}, but with a different formal representation.
While Dufourd uses a list of darts and links between these darts, Gonthier has a  more
generic way to represent finite sets.

The computation of convex hulls was also studied Meikle and Fleuriot, with the
focus on using Hoare logic to support the reasoning framework \cite{MeikleFleuriot2006}
and by Immler in the case of zonotopes, with applications
to the automatic proof of formulas \cite{DBLP:conf/cpp/Immler15}.

The algorithm we describe here is essentially the first phase of the one described in sections 3--4 of Lawson's report \cite{Lawson77}.

In the current state of our development, we benefit from the description of finite sets and
finite groups provided by the mathematical components library \cite{DBLP:journals/jfrea/GonthierM10,mcb}.  This library was initially
used for the four colour theorem \cite{Gonthier4CT} and further developed for the proof of the Feit-Thompson theorem \cite{DBLP:conf/itp/GonthierAABCGRMOBPRSTT13}.

Because it deals with the relative positions of points on a sphere, it is probable that the
Flyspeck formal development also contains many of the ingredients necessary to formalize triangulations \cite{Hales_etal._2017}.  For instance, Hales published a proof of the Jordan Curve theorem \cite{DBLP:journals/tamm/Hales07} that has
many similarities with the study of convex hulls and subdivisions of the plane.
\section{Conclusion}

The formal proofs described in this abstract have been developed with the Coq system \cite{Coq-8.8} and the mathematical components library \cite{mcb} and are available from
\begin{center}
\url{https://gitlab.inria.fr/bertot/triangles}
\end{center}
This is a preliminary study of the problem of building triangulations for a variety of purposes.  The naive algorithm is unsatisfactory as it does not provide a good way to find
the triangle inside of which a new point may occur.  This can be improved by using
Delaunay triangulations, as already studied formally in \cite{DBLP:conf/itp/DufourdB10} and a well-known
algorithm of ``visibility'' walk in the triangulation \cite{DevillersPT01}, which can be proved to have guarantees to terminate
only when the triangulation satisfies the Delaunay criterion.  This is the planned future work.

Delaunay triangulations, and their dual Vorono\"{\i} diagrams can be useful for practical problems concerning the motion of a device on a plane.  It will be useful to extend this work to
three dimensions and of course there already exists triangulation algorithms in three dimensions.  At first sight, the naive algorithm described here can be used directly for arbitrary dimensions, as long the notion of separating facet is given a suitable definition.  However, it seems that the proof done for the 2-dimensional case does not carry directly to a higher dimension \(d\): the boundary facets do not form a loop but a closed hyper-surface (of dimension \(d-1\)), there is not just a pair of purple points but a collection of purple facets of 
dimension \(d-2\).  Still some properties are preserved: the red facets are contiguous, and there are probably equivalents to Knuth's axioms for the higher dimensions.

\bibliography{main}
\bibliographystyle{splncs04}
\end{document}